\documentclass[12pt]{iopart}

\usepackage{amssymb}
\usepackage{psfrag}
\usepackage{graphicx}
\begin{document}

\title[]{$Q$-deformed description of excitons and associated physical results}

\author{M. Bagheri Harouni, R. Roknizadeh, M. H. Naderi}

\address{Quantum Optics Group,
Physics Department, University of Isfahan, Iran}
\ead{\mailto{m-baghreri@phys.ui.ac.ir},\
\mailto{rokni@sci.ui.ac.ir},\ \mailto{mhnaderi@phys.ui.ac.ir}}
\begin{abstract}
We consider excitons in a quantum dot as
$q$-deformed systems. Interaction of some excitonic systems with
one cavity mode is considered. Dynamics of the system is obtained
by diagonalizing total Hamiltonian and emission spectrum of
quantum dot is derived. Physical consequences of $q$-deformed
exciton on emission spectrum of quantum dot is given. It is shown
that when the exciton system deviates from Bose statistics,
emission spectra will become multi peak. With our investigation 
we try to find the origin of the $q$-deformation of exciton. The optical response of
excitons, which affected by the nonlinear nature of $q$-deformed
systems, up to the second order of approximation is calculated and
absorption spectra of the system is given.
\end{abstract}

\pacs{73.20.La,\;71.35.Cc,\;03.65.Fd}
\submitto{\JPB}

\maketitle
\section{Introduction}
Exciton is an elementary excitation of a semiconductor which
consists of a pair of two correlated fermions, the electron and
the hole. Analogous to the Hydrogen atom, it is characterized by a
binding energy $E_b$ and a Bohr radius $a_B$. Because an exciton
is composed of two fermions, it is a composite boson.
Particularly, in a bulk semiconductor, when the excitation of
system is dilute, i.e. $n_{ex}a_B^3\ll 1$, where $n_{ex}$ is the
exciton density, a bosonic description of system is convenient
\cite{davydov}. Also Bose-Einstein condensation of excitons, which
is an essential characteristic of boson systems has been
considered theoretically \cite{hanamura}. When density of excitons
increase, the above condition is violated. In this situation the
statistics of excitons deviates from Bose statistics.\\ \indent In
low dimensional semiconductor systems such as quantum well (QW),
quantum wire and quantum dot (QD), due to the small dimensions and
loss of translational symmetry, exciton excitation differs from
exciton in bulk materials. In semiconductor nanostructures the
size of the system strongly affects exciton properties. For
example, in the case of quantum well it is shown \cite{yamamoto1},
if the well width is larger (smaller) than the Bohr radius of
exciton, the spectrum of quantum well has properties similar to
the situation in which excitons are boson (fermion). Hence, the
size of the system directly affects the quantum statistics of
excitons in that system. Recently similar results has been
obtained for QD \cite{laussy}. In Ref.\cite{laussy} the
effects of different statistics of excitons on emission
spectra of a QD is investigated, and the origin of different statistics of excitons is
considered. The same results have also been obtained for the
quantum well. If the size of QD is smaller (larger) than the
exciton Bohr radius, excitons behave like fermion (boson). Real
statistics of excitons in the interaction is considered in
\cite{laikh} and references therein. As mentioned before in high
density regime, exciton statistics deviates from Bose statistics.
This is due to the increase of mutual forces between the excitations of
the system and then the Pauli exclusion principle plays a dominant
role \cite{combescot}.
 Appearance of
Bose statistics of exciton-biexciton system and Pauli exclusion
effects in superlattice has been considered
experimentally \cite{ichida}.\\
\indent Bosons and fermions are the only two kinds of particles
 realized in nature. The conditions mentioned for excitons (in
one regime they are like bosons and in another one like fermions)
are property of a special kind of statistics called intermediate
statistics \cite{khare}. Bose and Fermi statistics are two
limiting cases of this statistics. Properties of this statistics
have been considered by many authors \cite{blenco}$-$\cite{shen}.
Operator realization of intermediate statistics is similar to
$q$-deformed operators \cite{chator}. Bosonic $q$-deformed
operators \cite{macfarlan} are a generalization of the Heisenberg
algebra obtained by introducing a deformation parameter $q$.
Deviation of this parameter from 1 shows deviation of algebra from
the Heisenberg algebra. It is shown that it is possible to
describe correlated fermion pairs with $q$-deformed bosons
\cite{ezaf}. Therefore it is reasonable to consider an exciton
system as a $q$-deformed system. We assume the creation and
annihilation operators of excitons obey a $q$-deformed algebra. A
$q$-deformed
description of Frenkel exciton has been considered recently \cite{liu}. \\
\indent The algebra generated by $q$-deformed operators are given
by
\begin{eqnarray}\label{yek}
[\hat b_q,\hat b_q^\dag]_q&=&\hat b_q\hat b_q^\dag-q^{-1}\hat
b_q^\dag\hat b_q=q^{\hat n},\\ \nonumber [\hat n, \hat b_q^\dag]&=&\hat
b_q^\dag,\hspace{1.5cm} [\hat n, \hat b_q]=-\hat b_q.
\end{eqnarray}
where $\hat n=\hat b^\dag\hat b$ is the usual particle number
operator. Representation of this algebra is given in
\cite{rideau}. In the case of excitons, $q$-parameter can depend
on excitation number and physical size of system.\\ \indent In
this paper we consider the interaction of light with a QD embedded
in a microcavity. By considering excitons in QD as $q$-deformed
bosons (case of $q$-deformed fermion is straightforward) we study
the emission spectrum of the system. As is clear, the commutator
(\ref{yek}) explicitly depends on the number of excitons. Hence,
this is a system in which light interacts with a nonlinear active medium.
 Therefore, we shall obtain the linear and nonlinear
response of a $q$-deformed exciton system. Knowledge of interaction of light
with a nonlinear medium ($q$-deformed excitons) and its optical response
 is important for the interpretation of experimental results such as \cite{xu}.
 On the other hand, we compared the obtained results with some experimental
 ones and in this manner we investigate the physical origin of $q$-deformation
of excitons. In section 2 we derive the spectrum of
QD when one exciton mode interacts with a single mode
cavity-field. In section 3 we consider the interaction of two
exciton modes with a single cavity mode. In section 4 the
nonlinear response of QD is derived up to second order of
approximation. Finally we summarize our conclusions in section 5.


\section{Model Hamiltonian}
We consider a QD embedded in a microcavity which interacts with a
single mode cavity-field. We assume the excitations in QD have an
intermediate statistics \cite{laussy}, and their creation and
annihilation operators obey $q$-deformed algebra. We can express
the $q$-deformed operators in terms of ordinary boson operators by
the following maps
\begin{equation}\label{do}
  \hat b_q=\hat b\sqrt{\frac{q^{\hat n}-q^{-\hat n}}{\hat
  n(q-q^{-1})}}, \hspace{1.5cm}\hat b_q^\dag=\sqrt{\frac{q^{\hat n}-q^{-\hat n}}{\hat
  n(q-q^{-1})}}\hat b^\dag,
\end{equation}
where $\hat b$ and $\hat b^\dag$ are the ordinary boson operators
and $\hat n=\hat  b^\dag\hat b$. Ordinary commutator of
$q$-deformed exciton operators is
\begin{equation}\label{seh}
  [\hat b_q,\hat b_q^\dag]=\frac{q}{q+1}[q^n+q^{-(n+1)}]\equiv k(\hat
  n).
\end{equation}
Deviation of this commutator from ordinary boson algebra ($[\hat
b,\hat b^\dag]=1$) relates to deviation of $q$-parameter from 1.
It is clear, this generalized commutator depends on the number of
excitations. It seems that by using this algebra we can consider
some nonlinear phenomena in the system related to the population
of excitons. For example, biexciton effects can be considered in
this manner as an effective approach. So that the deformation
parameter $q$ can represent some physical parameters such as the
ratio of size of system to the Bohr radius of exciton. Interaction
of QD with single mode cavity-field in rotating wave approximation
can be described by the following Hamiltonian
\begin{equation}\label{char}
  \hat H=\hbar\omega\hat a^\dag\hat a+\hbar\omega_{ex}\hat
  b_q^\dag\hat b_q+\hbar g(\hat a\hat b_q^\dag+\hat a^\dag\hat
  b_q),
\end{equation}
where $\hat a$ and $\hat a^\dag$ are creation and annihilation
operators of cavity field and $[\hat a_i,\hat
a_j^\dag]=\delta_{ij}$. We shall consider a phenomenological
damping for the system which relates to both subsystems: photon
and exciton. As is clear from the Hamiltonian (\ref{char}), the
exciton number is not a constant of motion. Because of the
dependence of exciton operator, $\hat b_q$, on the exciton number,
resulting equations of motion become a nontrivial set of coupled
equations. On the other hand, since the total number of excitation
(exciton and photon) is conserved we can diagonalize the
Hamiltonian in the subspace of a definite excitation. To consider
this dynamics we propose an approach based on diagonalization of
the Hamiltonian by using the polariton transformation \cite{hop}. This
procedure depends on some unitary transformations which
diagonalize the model Hamiltonian. As is usual in this procedure
\cite{fano}, new operators have the same commutation relation as
the original operators (free operators). Here, there are two
distinct sets of operators, the cavity mode operators which obey
the usual boson commutation relation and exciton operators that
are $q$-deformed boson. Therefore, with the presence of these two
different statistics, mixed operators (polariton operators) do not
have specific statistics. They can be considered as ordinary boson
operators or $q$-deformed operators. We consider both situations
and we study the physical results associated with each situation
in the resonance fluorescence spectrum of QD.


\subsection{Boson polaritons} 
\begin{figure}
\begin{center}
\psfragscanon
\psfrag{W}[][][1.5]{$\omega\;(ev)$}
\psfrag{RL}[][][1.5]{$s(\omega)$}
\includegraphics[angle=0,width=.5\textwidth]{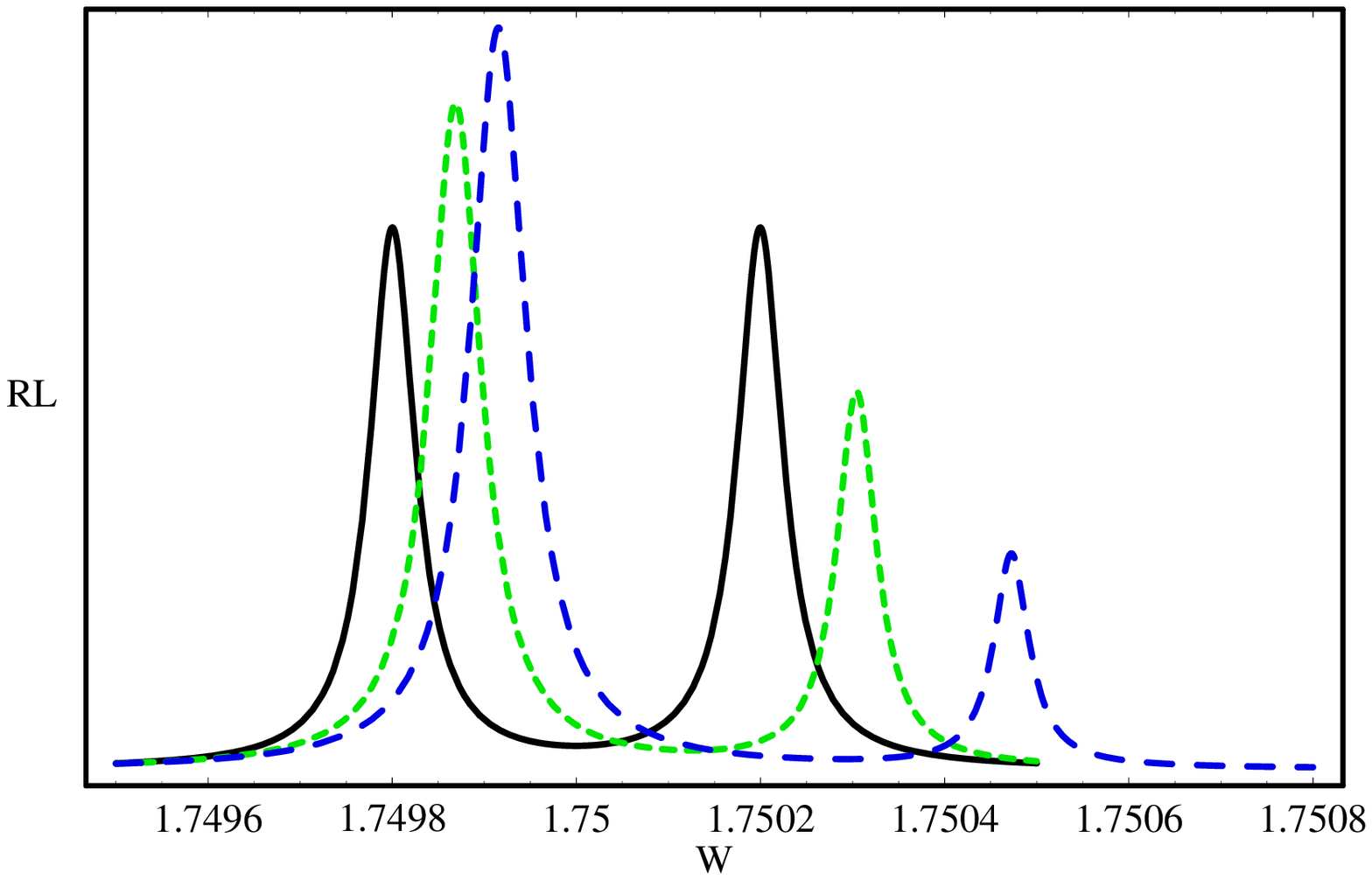}
 \caption{Plots of $S(\omega)$ versus $\omega$. Parameter are choose as $\omega=1.75\; eV$, $\omega_{ex}=1.75\; eV$,
  $g=200\;\mu eV$, $\gamma_{ex}=20\;\mu eV$, $\gamma_{ph}=40\;\mu eV$, $n=1$ and $|\alpha|^2=9$. Solid plot corresponds to $q=1$, nondeformed case.
 Dotted one corresponds to $q=1.01$, and for dash line $q$ is equal $1.015$.} \label{1}
\end{center}
\end{figure}

In order to solve the dynamical
system, we perform the following transformation
\begin{equation}\label{polariton}
  \hat p_k=u_k\hat b_q+v_k\hat a.
\end{equation}
Due to the presence of $q$-deformed operator $\hat b_q$, we call
this transformation a polariton-like transformation. As mentioned
before, $\hat b_q$ depends on the number of excitons explicitly
and this causes the Hopfield coefficients $u_k$ and $v_k$ will
depend on the number of excitons. Hence, the transformation
(\ref{polariton}) can be considered as a nonlinear polariton
transformation. This kind of transformation has been considered
recently for the case of Bogoliubov transformation
\cite{katriea,naderi}. We assume polariton-like operators obey the
usual boson commutation relations
\begin{equation}\label{}
  [\hat p_k,\hat p^\dag_{k'}]=\delta_{kk'}\Rightarrow[\hat p_k,\hat
  p^\dag_{k}]=|u_k|^2k(\hat n)+|v_k|^2=1,
\end{equation}
where the operator valued function $k(\hat n)$ was introduced by
Eq.(\ref{seh}). We choose unknown coefficients $u_k$ and $v_k$ so
that the Hamiltonian (\ref{char}) becomes diagonal in terms of the
polariton-like operators
\begin{equation}\label{diah}
  \hat H=\hbar\sum_k\Omega_k\hat p_k^\dag\hat p_k,
\end{equation}
where $\Omega_k$ is the polariton spectrum and $k$ refers to
different polariton branches. By taking into account a
phenomenological damping for exciton and photon systems
separately, the unknown parameters satisfy following set of
equations
\begin{equation}\label{set1}
  [\omega_{ex}k(\hat n)-\Omega_k-i\gamma_{ex}]u_k+v_kg=0, \hspace{1.1cm}
  u_kgk(\hat n)+(\omega-\Omega_k-i\gamma_{ph})v_k=0.
\end{equation}
In this set of equations, $\gamma_{ex}$ and $\gamma_{ph}$ are the
exciton and photon damping constants, respectively. From these
equations the polariton spectrum can be obtained as
\begin{eqnarray}\label{spect}
  \Omega_k&=&\frac{\omega_{ex}k(\hat n)+\omega-i(\gamma_{ex}+\gamma_{ph})}{2}\nonumber \\ &\pm&
  \frac{1}{2}\sqrt{[\omega_{ex}
  k(\hat n)-\omega-i(\gamma_{ex}-\gamma_{ph})]^2+4g^2k(\hat n))}.
\end{eqnarray}
It is apparent that $q$-deformed description of excitons causes
the splitting between these energy eigenvalues be increased in
compare to the case of bosonic description of exciton. Using the set
of equations (\ref{set1}) and the polariton spectrum (\ref{spect})
we find the coefficients for two polariton branches
\begin{eqnarray}\label{coeff1}
 && u_k=\sqrt{\frac{\omega-i\gamma_{ph}-\Omega_k}{k(\hat n)[\omega-2\Omega_k+\omega_{ex}k(\hat
  n)-i(\gamma_{ex}+\gamma_{ph})]}},\\ \nonumber &&v_k=-\sqrt{\frac{\omega_{ex}k(\hat n)-i\gamma_{ex}-\Omega_k}
  {\omega-2\Omega_k+\omega_{ex}k(\hat n)-i(\gamma_{ex}+\gamma_{ph})}}.
\end{eqnarray}
By employing these coefficients all necessary parameters for the
polariton Hamiltonian are determined.
\\ \indent Now we can consider the dynamics of polariton operators. The time evolution of polariton operators is governed
by the polariton Hamiltonian (\ref{diah})
\begin{equation}\label{}
  \hat{\dot{p}}_k=\frac{-i}{\hbar}[\hat p_k, \hat H]=-i\Omega_k\hat
  p_k.
\end{equation}
Let us consider damping effects by taking into account a
phenomenological damping term and noise operator in the dynamical
equations of polariton operators. Hence, the time evolution of
polariton operator is given by
\begin{equation}\label{poperat}
\hat{\dot{p}}_k=-i\Omega_k\hat p_k-\Gamma_k\hat p_k+\hat F_{\hat
p_k}(t),
\end{equation}
where $\hat F_{\hat p_k}(t)$ is the Langevin noise operator which
depends on the reservoir variables and $\Gamma_k$ is the damping
constant of $k$th polariton branch given by
$\Gamma_k=\frac{\gamma_{ex}+\gamma_{ph}}{2}$. Correlation
functions of the noise operators determine physical properties of
the system. The Langevin noise operator are such that their
expectation values $\langle \hat F_x\rangle$ vanishes, but their
second order moments do not \cite{lax}. They are intimately linked
up with the global dissipation and in a Markovian environment they
take the form
\begin{equation}\label{}
  \langle \hat F^\dag_{\hat p_k}(t)\hat F_{\hat
  p_k}(t')\rangle=2\Gamma_k\delta(t-t').
\end{equation}
With neglecting the phonon effects by decreasing the temperature,
other sources of damping like spontaneous recombination of exciton and photon loss
 are considered as Markovian procedures. It follows, on solving
Eq.(\ref{poperat}), that
\begin{equation}\label{potim}
 \hat{p}_k(t)=\hat{p}_k(0)e^{(-i\Omega_k-\Gamma_k)t}+\int_0^te^{(-i\Omega_k-\Gamma_k)(t-t')}\hat F_{\hat
 p_k}(t')dt'.
\end{equation}
In this equation we set initial time equal zero. \\ \indent The
power spectrum of the scattered light for statistical stationary
fields is given by \cite{scully}
\begin{equation}\label{}
  S(r,\omega)=\frac{1}{\pi}Re\int_0^{\infty}\langle\hat
  E^-(r,t)\hat E^+(r,t+\tau)\rangle e^{i\omega \tau}d\tau,
\end{equation}
where $\hat E^\pm$ are the positive and negative frequency parts
of the electric field operator. Expressing field operators in
terms of creation and annihilation operators we have
\begin{equation}\label{}
  S(r,\omega)=\frac{A(r)}{\pi}Re\int_0^{\infty}\langle\hat a^\dag(0)\hat
  a(\tau)\rangle e^{i\omega\tau}d\tau.
\end{equation}
Here, we set $t=0$, and $A(r)$ depends on mode function of the
cavity-field.\\ \indent Now we can express, the field and exciton
creation and annihilation operators in terms of polariton ones:
\begin{equation}\label{}
  \hat a=v_1^\ast\hat p_1+v_2^\ast\hat p_2, \hspace{1.5cm} \hat
  b_q=k(\hat n)(u_1^\ast\hat p_1+u_2^\ast\hat p_2),
\end{equation}
and at the time $t$ we have
\begin{equation}\label{}
  \hat a(t)=v_1^\ast\hat p_1(t)+v_2^\ast\hat p_2(t).
\end{equation}
Now to calculate the resonance fluorescence spectrum we have to
determine the initial state of system. we assume at $t=0$, the
cavity-field is in a coherent state $|\alpha\rangle$, and the
exciton subsystem in its vacuum state. Under this condition, by
using Eq.(\ref{potim}) the resonance fluorescence spectrum is 
obtained as
\begin{equation}\label{}
  S(r,\omega)=\frac{A(r)|\alpha|^2}{\pi}\left[|v_1|^2\frac{\Gamma_1}{(\omega-\Omega_1)^2+\Gamma_1^2}+
  |v_2|^2\frac{\Gamma_2}{(\omega-\Omega_2)^2+\Gamma_2^2}\right].
\end{equation}
In deriving this result we implicitly assume that at $t=0$ the
noise operator and polariton operators are uncorrelated.
Fig.(\ref{1}) shows the plot of $S(r,\omega)$ versus $\omega$ for
different values of deformation parameter q. Material parameters
 are chosen  as $\omega=1.75\;eV$, $\omega_{ex}=1.75\;eV$,
  $g=200\;\mu eV$, $\gamma_{ex}=20\;\mu eV$, $\gamma_{ph}=40\;\mu eV$ \cite{peter}, $n=100$ and $|\alpha|^2=9$
. As is clear when $q=1$, spectrum has similar
variation as experimental results \cite{peter}. This figure shows
that when $q=1$ (nondeformed case) the power spectrum of the
fluorescence light is a double peak centered at $\omega=\Omega_1$
and $\omega=\Omega_2$. By increasing deviation of q from 1, it is
apparent from the different plots in this figure that splitting
between two peaks increases and the height of one of peaks
decreases. This result has been reported in resonance fluorescence
of excitons when the biexcitonic interaction is taken into
account. It has been shown \cite{yamamoto1} that biexcitonic
effects are a red shift of the transition frequencies, emergence
of sidebands due to the switch-on forbidden transitions and
asymmetry of the emission spectrum. The binding energy of biexciton in QD causes a shift
 in the spectrum of the system. In the present model the splitting of spectrum (Rabi splitting)
 depends on the $q$-parameter. Hence, changing this parameter affects the spectrum. Then as a one reason of deviation of excitons
 from ideal Bose system  we can consider Coulomb interaction between them.
 On the other hand, $q$-deformed exciton operators depend on
the total number of exciton, and biexciton interaction occurs when
there are more than one exciton. This similarity makes this clue
that the $q$-deformation can be consider as an effective approach
to take into account the biexciton effects. As mentioned before,
the $q$-parameter can depend on the size of sample. The plotted
resonance fluorescence spectrum in Fig.(\ref{1}) makes clear some
differences of optical properties of different size QD. For large
values of q, compare with 1, spectrum will reduce to one peak.
This case is a characteristic of the weak coupling regime.


\subsection{$Q$-deformed polaritons}
\begin{figure}
\begin{center}
\psfragscanon
\psfrag{W}[][][1.5]{$\omega\;(ev)$}
\psfrag{RL}[][][1.5]{$s(\omega)$}
\includegraphics[angle=0,width=.5\textwidth]{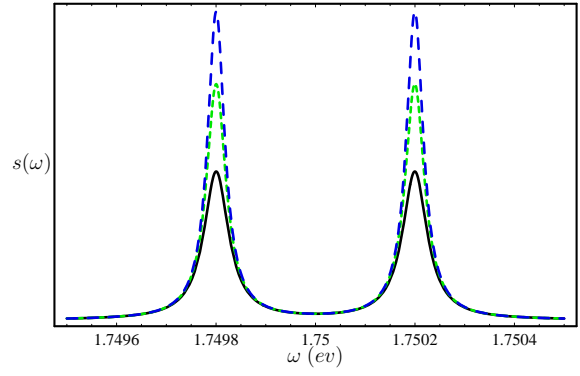}
 \caption{Plots of $S(\omega)$ versus $\omega$. Parameter are choose as $\omega=1.75\; eV$, $\omega_{ex}=1.75 \;eV$,
  $g=200\;\mu eV$, $\gamma_{ex}=20\;\mu eV$, $\gamma_{ph}=40\;\mu eV$, $n=1$ and $|\alpha|^2=9$. In all figures we have $q=1$.
  Solid line
   corresponds to case $s=1$. In dotted one we have $s=1.007$ and for dash line $s=1.01$.} \label{2}
\end{center}
\end{figure}

In this subsection we assume that the polariton operators are
$q$-deformed operators. According to the $q$-deformed nature of
the exciton system we assume the following algebra for polariton
operators
\begin{equation}\label{}
  [\hat p_k,\hat p^\dag_k]_s=\hat p_k\hat p^\dag_k-s^{-1}\hat p^\dag_k\hat
  p_k=s^{\hat n_k},
\end{equation}
where $s$ denotes the deformation parameter corresponding to the
polariton system and $\hat n_k$ shows the number operator for $
k$th polariton branch. Ordinary commutator for these operators is
\begin{equation}\label{}
  [\hat p_k,\hat p^\dag_k]=|u_k|^2k(\hat
  n)+|v_k|^2=\frac{s}{s+1}[s^{\hat n_k}+s^{-(\hat n_k+1)}]=M(\hat n_k).
\end{equation}
Using the same approach of the previous subsection we obtain the
following set of equations for the coefficients of transformation
\begin{eqnarray}\label{}
  [(\omega_{ex}k(\hat
  n)-i\gamma_{ex}-\Omega'_kM(n_k)]u_k+v_kg&=&0,\nonumber \\u_kgk(\hat
  n)+[\omega-i\gamma_{ph}-\Omega'_kM(n_k)]v_k&=&0.
\end{eqnarray}
From this set of equations we derive the deformed polariton
spectrum as
\begin{eqnarray}\label{}
  \Omega'_k&=&\frac{\omega_{ex}k(\hat n)+\omega-i(\gamma_{ex}+\gamma_{ph})}{2M(n_k)}\nonumber \\&\pm&
  \frac{\sqrt{[\omega_{ex}
  k(\hat n)-\omega-i(\gamma_{ex}-\gamma_{ph})]^2+4g^2k(\hat n))}}{2M(n_k)}.
\end{eqnarray}
and the transformation coefficients read as
\begin{eqnarray}\label{}
  && u_k=-\sqrt{\frac{M(n_k)[\omega-i\gamma_{ph}-\Omega'_kM(n_k)]}{k(\hat n)[\omega-2\Omega'_kM(n_k)+\omega_{ex}k(\hat
  n)-i(\gamma_{ex}+\gamma_{ph})]}}, \\ \nonumber && v_k=\sqrt{\frac{M(n_k)[\omega_{ex}k(\hat n)-i\gamma_{ex}-\Omega'_kM(n_k)]}{\omega-2\Omega'_kM(n_k)+\omega_{ex}k(\hat
  n)-i(\gamma_{ex}+\gamma_{ph})}}.
\end{eqnarray}
By determining all the variables, polariton Hamiltonian (diagonal
Hamiltonian) will be determined. By applying the same procedure as
before we derive the resonance fluorescence spectrum in this case
as follows
\begin{equation}\label{}
  S(r,\omega)=\frac{A(r)|\alpha|^2(|v_1|^2+|v_2|^2)}{\pi}\sum_{i=1,2}|v_i|^2\frac{\Gamma_i}
  {(\omega-\Omega'_iM(n_k))^2+\Gamma_i^2}.
\end{equation}
 Fig. (\ref{2}) shows the plot of $S(r,\omega)$ versus $\omega$ for
different values of polariton deformation parameter $s$. This
figure shows that changes of s-parameter (deformation parameter of
polariton) does not cause any shift in transition frequencies,
but causes strengths of peaks increase.


\section{Interaction of light with two exciton modes}
\begin{figure}
\begin{center}
\psfragscanon
\psfrag{W}[][][1.5]{$\omega\;(ev)$}
\psfrag{RL}[][][1.5]{$s(\omega)$}
\includegraphics[angle=0,width=.5\textwidth]{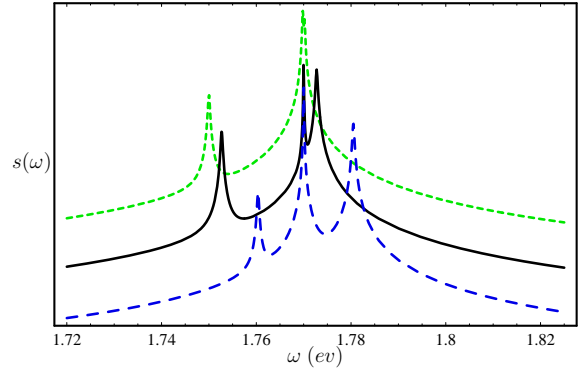}
 \caption{Plots of $S(\omega)$ versus $\omega$. Parameter are choose as $\omega=1.75\;eV$, $\omega_{ex_1}=1.75\; eV$,
  $\omega_{ex_2}=1.77\;eV$, $g=200\;\mu eV$, $\gamma_{ex_1}=\gamma_{ex_2}=200\;\mu eV$, $\gamma_{ph}=45\;\mu eV$,
   $n_1=1$, $n_2=1$ and $|\alpha|^2=9$. Dotted line corresponds to nondeformed case $q_1,q_2=1$. For solid line $q_1,q_2=1.04$.
   In the case of dashed line $q_1,q_2=1.08$.} \label{3}
\end{center}
\end{figure}

We now consider the interaction of one cavity mode with QD when
two exciton modes are coupled to the field mode. As before, we
assume exciton system is expressed by the $q$-deformed operators.
The total Hamiltonian of the system under consideration can be
written as follows
\begin{equation}\label{hamise}
  \hat H=\hbar\omega\hat a^\dag\hat
  a+\hbar\sum_{i=1,2}\omega_{{ex}_i}\hat b_{q_i}^\dag\hat b_{q_i}+\hbar
  g\sum_{i=1,2}(\hat a\hat b_{q_i}^\dag+\hat a^\dag\hat b_{q_i}).
\end{equation}
We assume both excitons have the same coupling constant with the
cavity mode. We solve this system as before by diagonalizing the
Hamiltonian. For this purpose we perform the following
transformation
\begin{equation}\label{polse}
  \hat p_k=u_k\hat b_{q_1}+x_k\hat b_{q_2}+v_k\hat a.
\end{equation}
We consider the situation in which the polariton operators obey
the nondeformed Bose statistics
\begin{equation}\label{}
  [\hat p_k,\hat p^\dag_k]=|u_k|^2k(\hat n_1)+|x_k|^2k(\hat n_2)+|v_1|^2=1,
\end{equation}
where $\hat n_i$ ($i=1,2$) represents the number operator for each
excitonic mode. As is clear in this case there are three polariton
branches. Assuming the transformation (\ref{polse}) diagonalizes
the Hamiltonian (\ref{hamise}), this polariton Hamiltonian takes
the following form
\begin{equation}\label{}
  \hat H=\hbar\sum_k\Omega_k\hat p^\dag_k\hat p_k,
\end{equation}
where summation is over all polariton branches. The following
equation determines the polariton spectrum
\begin{equation}\label{}
  (c-\Omega_k)[(d-\Omega_k)(\omega-i\gamma_{ph}-\Omega_k)-g^2k(\hat n_1)]-g^2k(\hat
  n_2)(d-\Omega_k)=0,
\end{equation}
where $c=\omega_{ex_1}k(\hat n_1)-i\gamma_{ex_1}$ and
$d=\omega_{ex_2}k(\hat n_2)-i\gamma_{ex_2}$. By deriving the
polariton spectrum the transformation parameters are obtained as
\begin{eqnarray}\label{}
 &&u_k=\frac{g[(d-\Omega_k)(\omega-i\gamma_{ph}-\Omega_k)-g^2k(\hat n_2)]}{A},
 \\ \nonumber &&x_k=\frac{g^3k(\hat n_1)}{A}\\ \nonumber
 &&v_k=-\frac{(c-\Omega_k)[(d-\Omega_k)(\omega-i\gamma_{ph}-\Omega_k)-g^2k(\hat
 n_2)]}{A},
\end{eqnarray}
where the parameter $A$ is given by
\begin{eqnarray}\label{}
  A&=&([g^2k(\hat n_1)+(c-\Omega_k)^2][(d-\Omega_k)(\omega-i\gamma_{ph}-\Omega_k)-g^2k(\hat n_2)]^2\nonumber \\&+&g^6k^2(\hat n_1)k(\hat
  n_2))^{\frac{1}{2}}.
\end{eqnarray}
In this manner, all the parameters which appear in the polariton
Hamiltonian are determined. By repeating the approach of previous
section the resonance fluorescence spectrum of system with
different initial conditions can be determined. If we assume at
$t=0$ the cavity mode is in the coherent state $|\alpha\rangle$
and QD in vacuum state $|0\rangle$, the resonance fluorescence
spectrum is given by
\begin{equation}\label{}
  S(r,\omega)=\frac{|\alpha|^2A(r)}{\pi}\sum_k\frac{|v_k|^2\Gamma_k}{\Gamma_k^2+(\omega-\Omega_k)^2}.
\end{equation}
To show complex structure (multi-peak structure) of this spectrum
Fig. (\ref{3}) presents the spectra on a logarithmic scale. For the sake of 
clarity, we have powered some peaks compare to other ones in this figure. In
the case of $q=1$ (nondeformed exciton) the spectrum has two
peaks. Increasing the $q$-parameter causes that splitting between peaks be increased and spectrum 
becomes multi-peaks. Multi-peaks structure in
emission of exciton such as Mollow triplet was predicted when
excitons obey statistics different from Bose statistics
\cite{yamamoto1,laussy}. When, $q$-parameter is changed, the
energy and intensities of emission change. Effects of exciton
number on absorption spectrum of QD is considered . Due to the
relation of absorption spectrum and resonance fluorescence,
similar result is obtain in \cite{franc}.


\section{Nonlinear response of excitons in $q$-deformed regime}
\begin{figure}
\begin{center}
\psfragscanon
\psfrag{W}[][][1.5]{$\omega\;(mev)$}
\psfrag{a}[][][1.5]{$\alpha(\omega)$}
\includegraphics[angle=0,width=.5\textwidth]{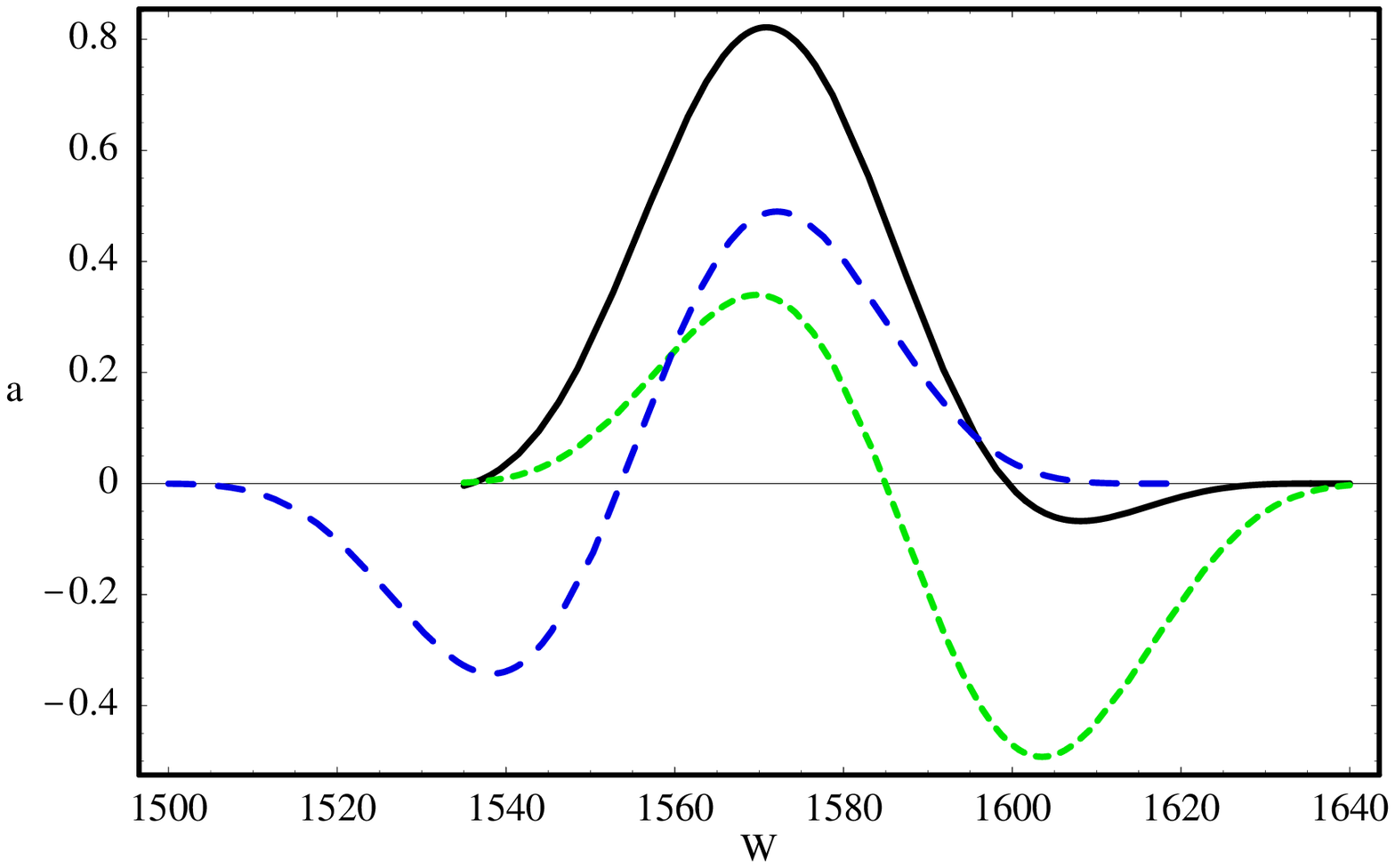}
 \caption{Plots of spectrum absorption versus $\omega$. We consider 1s-exciton and Parameter
  are choose as $\hbar=e=1$, $g=200\;\mu ev$ and $\omega_{ex}=1574 \;mev$.
  Solid plot corresponds to nondeformed case $q=1$. For dotted one
  $q=1.01$ and
   in dash one $q=0.99$.} \label{4}
\end{center}
\end{figure}

\begin{figure}
\begin{center}
\psfragscanon
\psfrag{W}[][][1.5]{$\omega\;(mev)$}
\psfrag{a}[][][1.5]{$\alpha(\omega)$}
\psfrag{q}[][][1.5]{q}
\includegraphics[angle=0,width=.5\textwidth]{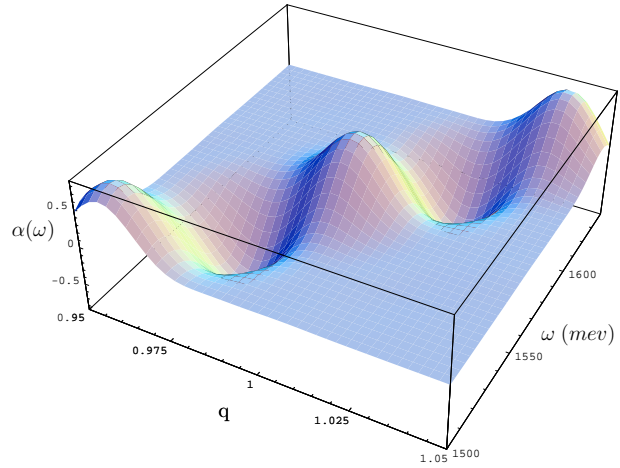}
 \caption{3D-Plots of spectrum absorption versus $\omega$ and deformation parameter $q$. Physical
 parameter are the same as Fig.(\ref{4}).} \label{5}
\end{center}
\end{figure}

\begin{figure}
\begin{center}
\psfragscanon
\psfrag{W}[][][1.5]{$\omega\;(mev)$}
\psfrag{a}[][][1.5]{$\alpha^{(3)}(3\omega)$}
\includegraphics[angle=0,width=.5\textwidth]{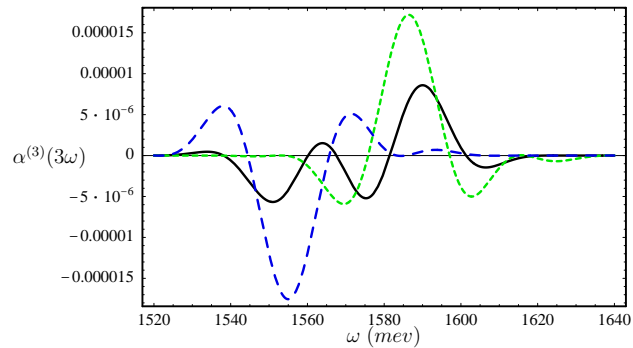}
 \caption{Plots of nonlinear spectrum absorption versus $\omega$. We consider 1s-exciton and
 Parameters
  are choose as $\hbar=e=1$, $g=200\;\mu ev$ and $\omega_{ex}=1574\; mev$
 Solid plot corresponds to nondeformed case $q=1$. In dotted plot $q=1.01$.
   In dash plot $q=0.99$.} \label{6}
\end{center}
\end{figure}

In previous sections we considered some physical results of
$q$-deformed description of excitons. The $q$-deformed description
can be served as a nonlinear description of excitons. It is
well-known that different kinds of nonlinearity in an exciton
system lead to different orders of nonlinear response of the
system \cite{axt,sham}. Therefore, we try to obtain optical
response of a driven quantum dot, which its optical excitations
are considered as $q$-deformed systems. For this purpose we will
calculate the coefficient absorption of a QD in this regime. In
this section we neglect all damping effects and we consider the
Hamiltonian of the system as follows
\begin{equation}\label{}
  \hat H=\hbar\omega\hat a^\dag\hat a+\hbar\omega_{ex}\hat
  b_q^\dag\hat b_q+\hbar g(\hat a\hat b_q^\dag+\hat a^\dag\hat
  b_q).
\end{equation}
In the electron picture, the induced dipole moment by transition
of an electron is described by $\hat \mu=\hat a^\dag_v\hat
a_c+\hat a^\dag_c\hat a_v$ \cite{koch}. The operator $\hat
a^\dag_v\;(\hat a_v)$ is the creation (annihilation) operator for
an electron in the valance band (level in the case of QD), and
$\hat a^\dag_c\;(\hat a_c)$ is the creation (annihilation)
operator for an electron in the conduction band. Hence, creation
of an exciton is denoted by $\hat a^\dag_c\hat a_v=\hat
b_q^\dag$. Therefore we can write the dipole operator of QD as
$\hat\mu=\hat b^\dag_q+\hat b_q$. The macroscopic polarization is
expectation value of polarization operator. The optical response
function represents the reaction of the system to an external
classic field $E(t)$ coupled to the variables of system
\cite{mukamel}, i.g., the dipole operator. Hence, we consider an
external field as a pump source and we treat the reaction of QD
to it. Then the total Hamiltonian of system is then given by
\begin{equation}\label{tothami}
\hat H=\hbar\omega\hat a^\dag\hat a+\hbar\omega_{ex}\hat
  b_q^\dag\hat b_q+\hbar g(\hat a\hat b_q^\dag+\hat a^\dag\hat
  b_q)-[\vec d_{vc}\cdot\vec E(t)\hat b_q+\vec d_{cv}\cdot\vec E(t)\hat
  b^\dag_q],
\end{equation}
where $\vec d_{vc}$ denotes the dipole matrix element. The
Hamiltonian in the interaction picture has the form
\begin{eqnarray}\label{hintk}
  \hat H_{int}&=&\hbar g\left[\hat a\hat b^\dag_qe^{-i[\omega-\omega_{ex}k(\hat n)]t}+
  \hat a^\dag e^{i[\omega-\omega_{ex}k(\hat n)]t}\hat
  b_q\right]\\ \nonumber &-&\left[\vec d_{cv}\cdot\vec E(t)e^{-i\omega_{ex}k(\hat n)t}\hat b_q+\vec d_{vc}\cdot\vec E(t)
  \hat b^\dag_qe^{i\omega_{ex}k(\hat n)t}\right].
\end{eqnarray}
The observable of interest for the optical response is the
time-dependent dipole density $\mu(t)=\langle\hat
b_q(t)\rangle+h.c.=Tr_{ex}(\hat b_q\rho_{ex}(t))+h.c.$, where
$Tr_{ex}$ means trace over the exciton system and
$\rho_{ex}(t)=Tr_f\rho(t)$, which $\rho(t)$ is the total time
dependent density matrix of the system and $\rho_{ex}(t)$ is the
time dependent density matrix of exciton system. The total time
dependent density matrix is given by
\begin{equation}\label{}
  \hat\rho(t)=\hat U(t,t_0)\hat \rho(t_0)\hat U^{-1}(t,t_0),
\end{equation}
where $U(t,t_0)=\hat T\exp[-\frac{i}{\hbar}\int_{t_0}^t\hat
H_{int}(t')dt']$ is the time ordered evolution operator and
$\hat\rho(t_0)$ is the total density matrix of system at initial
time. We assume that the quantum field and exciton system are
both in vacuum state. Therefore, the time dependent density matrix
of excitons is given by
\begin{equation}\label{}
  \hat\rho_{ex}(t)=\sum_n\langle n|\hat U(t,t_0)(|0\rangle_f|0\rangle_{ex})(_{ex}\langle0|_f\langle
  0|)\hat U^{-1}(t,t_0)|n\rangle,
\end{equation}
where summation is carried on field state and the matrix elements
of the time evolution operator are in the basis of field states.
By using the Feynman disentanglement theorem \cite{feynman} the
matrix elements of the time evolution operator $\hat U(t,t_0)$ can
be evaluated. We can write Hamiltonian in (\ref{hintk}) as $\hat
H_{int}=\hat H_1(t)+\hat H_2(t)$, where
\begin{eqnarray}\label{}
  &&\hat H_1(t)=\hbar g\left[\hat a\hat b^\dag_qe^{-i[\omega-\omega_{ex}k(\hat n)]t}+
  \hat a^\dag e^{i[\omega-\omega_{ex}k(\hat n)]t}\hat
  b_q\right]\\ \nonumber
  &&\hat H_2(t)=-\left[\vec d_{cv}\cdot\vec E(t)e^{-i\omega_{ex}k(\hat n)t}\hat b_q+\vec d_{vc}\cdot\vec E(t)
  \hat b^\dag_qe^{i\omega_{ex}k(\hat n)t}\right].
\end{eqnarray}
As is clear $\hat H_2(t)$ depends only on exciton operators. The
time evolution operator can be written as
\begin{eqnarray}\label{}
  \hat U(t,t_0)&=&\hat T\exp[-\frac{i}{\hbar}\int_{t_0}^t(\hat H_1(t')+\hat
  H_2(t))dt']\nonumber \\&=&\hat T\exp[-\frac{i}{\hbar}\int_{t_0}^t\hat H_2(t')dt']\exp[-\frac{i}{\hbar}\int_{t_0}^t\hat
  H_1(s)ds].
\end{eqnarray}
In this equation we use Feynman notation \cite{feynman}. These two
exponential terms are not disentangle from each other. They are
correlated and in doing integration, we have to take into account
ordering of operators. In calculation of matrix element of this
operator in the basis of field states, second exponential can be
considered as a ordinary c-number function of $t'$, because it is
independent of field operators:
\begin{equation}\label{}
  \langle i|\hat U(t,t_0)|j\rangle=\langle i|\hat T\exp[-\frac{i}{\hbar}\int_{t_0}^t\hat
  H_1(t')dt']|j\rangle\exp[-\frac{i}{\hbar}\int_{t_0}^t\hat H_2(s)ds].
\end{equation}
On the other hand we consider all the exciton operators in $\hat
H_1(t)$ as ordinary c-number functions, and we can write
\begin{eqnarray}\label{efefefe}
  \langle i|\hat U(t,t_0)|j\rangle&=&\langle i|\hat T\exp[-\frac{i}{\hbar}\int_{t_0}^t(\hat a^\dag_{t'}e^{i\omega t'}B(t')
  +\hat a_{t'}e^{-i\omega t'}B^\ast(t'))dt']|j\rangle\nonumber\\&\times&\exp[-\frac{i}{\hbar}\int_{t_0}^t\hat
  H_2(s)ds],
\end{eqnarray}
where $B(t)$ is a ordinary function corresponding to exciton
operators. As is clear this matrix element is a function of
exciton operators. The influence of exciton system is completely
contained in this operator functional and factored term in
(\ref{efefefe}). By using Feynman theorem, above matrix element
can written as
\begin{equation}\label{}
  \langle i|\exp[-ig\int_{t_0}^t\hat a^\dag_{t'}e^{i\omega t'}B(t')dt']\exp[-ig\int_{t_0}^t\hat a'_{t'}e^{-i\omega
  t'}B^\ast(t')dt']|j\rangle,
\end{equation}
where in this equation $\hat a'_{t'}=\hat V^{-1}(t)\hat a_t\hat
V(t)$, and
\begin{equation}\label{}
  \hat V(t)=\exp[-ig\hat a^\dag\int_{t_0}^tB(t')e^{i\omega t'}].
\end{equation}
In this manner the density matrix of exciton system takes the
following form
\begin{equation}\label{final}
  \hat\rho_{ex}(t)=\sum_n\frac{1}{n!}S_1(\hat n, \hat b_q, \hat
  b^\dag_q)|0\rangle_{ex}\langle0|S_2(\hat n, \hat b_q, \hat
  b^\dag_q),
\end{equation}
where
\begin{eqnarray}\label{}
S_1(\hat n, \hat b_q, \hat b^\dag_q)&=&\left[-g\hat
b^\dag_q\frac{e^{i\omega_{ex}k(\hat
n_{ex})(t-t_0)}}{\omega_{ex}k(\hat
n_{ex})}\right]^ne^{-\frac{g^2}{2}L(\hat n_{ex})}f(\hat b_q,\hat
b_q^\dag),\nonumber \\ S_2(\hat n, \hat b_q, \hat b^\dag_q)&=&\left[-g\hat
b_q\frac{e^{-i\omega_{ex}k(\hat
n_{ex}+1)(t-t_0)}}{\omega_{ex}k(\hat
n_{ex}+1)}\right]^ne^{-\frac{g^2}{2}L(\hat n_{ex})}f^{-1}(\hat
b_q,\hat b_q^\dag),\nonumber \\ L(\hat n_{ex})&=&\hat b_q^\dag\hat
b_q[\frac{e^{-i\omega_{ex}[k(\hat n_{ex}+1)-k(\hat
n_{ex}-1)](t-t_0)}}{[\omega-\omega_{ex}k(\hat
n_{ex}-1)][\omega-\omega_{ex}k(\hat n_{ex}+1)]} \\ 
&+&\frac{e^{-i\omega_{ex}[k(\hat n_{ex}+2)-k(\hat n_{ex})
)](t-t_0)}}{[\omega-\omega_{ex}k(\hat
n_{ex}+2)][\omega-\omega_{ex}k(\hat n_{ex})]}],
\end{eqnarray}
and
\begin{eqnarray}\label{}
f(\hat b_q, \hat b_q^\dag)=\hat
T\exp&[&\frac{i}{\hbar}\int_{t_0}^tdt'(\vec
d_{cv}\cdot\vec E(t')\hat b_qe^{-i\omega_{ex}k(\hat
n_{ex}+1)t'}\nonumber \\&+&\vec d_{vc}\cdot\vec E(t')
  \hat b^\dag_qe^{i\omega_{ex}k(\hat n_{ex})t'})],
\end{eqnarray}
By expanding the function $f(\hat b_q,\hat b^\dag_q)$ up to
second order in $E(t)$ and using (\ref{final}) we obtain the
time-dependent dipole density as follows
\begin{eqnarray}\label{}
&&p(t)=\sum_n\frac{g^{2n}}{n!}h_1(n)!\sqrt{f_q(n)!}e^{-\frac{g^2}{2}L(n)}\times\nonumber\\
 &&[\frac{i}{\hbar}h_0(n+1)!e^{-\frac{g^2}{2}L(1)}
\sqrt{f_q(n+1)!f_q(n+1)}\int_{t_0}^t\vec d_{cv}\cdot\vec
E(t')e^{i\omega_{ex}t'}dt'\nonumber \\&&
-\frac{i}{\hbar}\sqrt{f_q(n)}h_0(n)!e^{-\frac{g^2}{2}L(0)}
\sqrt{f_q(n)!f_q(n)}\int_{t_0}^t\vec d_{cv}\cdot\vec
E(t')e^{i\omega_{ex}k(n-1)t'}dt'\nonumber \\&&
+\frac{i}{2\hbar^3}\sqrt{f_q(n+1)}h_0(n+2)!e^{-\frac{g^2}{2}L(0)}
\sqrt{f_q(n+2)!f_q(n+2)}\times \nonumber \\&&\int_{t_0}^t\int_{t_0}^t\int_{t_0}^t\vec
d_{vc}\cdot\vec E(t')\vec d_{cv}\cdot\vec
E(r)\vec d_{cv}\cdot\vec
E(s)e^{i\omega_{ex}[s+r-k(n+1)t']}dt'drds \nonumber \\&&+\frac{i}{2\hbar^3}\sqrt{f_q(n)}h_0(n)!e^{-\frac{g^2}{2}L(0)}
\sqrt{f_q(n)!f_q(n)}\times \nonumber\\
&&\int_{t_0}^t\int_{t_0}^t\int_{t_0}^t\vec
d_{cv}\cdot\vec E(t')\vec d_{vc}\cdot\vec E(r)\vec d_{cv}\cdot\vec
E(s)e^{i\omega_{ex}[k(n-1)t'-(r-s)]}dt'drds \nonumber \\
&&-\frac{i}{2\hbar^3}\sqrt{f_q(n+1)}h_0(n+1)!e^{-\frac{g^2}{2}L(1)}
\sqrt{f_q(n+1)!f_q(n+1)}\int_{t_0}^t\int_{t_0}^t\int_{t_0}^t\times
\nonumber \\&&\vec d_{vc}\cdot\vec E(r)\vec d_{cv}\cdot\vec E(s)\vec
d_{cv}\cdot\vec E(t')e^{-i\omega_{ex}[k(n+1)(r-s)-t']}dt'drds 
\nonumber\\
&&-\frac{i}{2\hbar^3}\sqrt{f_q(n)}h_0(n+1)!e^{-\frac{g^2}{2}L(1)}
\sqrt{f_q(n+1)!f_q(n+1)}\int_{t_0}^t\int_{t_0}^t\int_{t_0}^t\times
\nonumber \\&& \vec d_{cv}\cdot\vec E(r)\vec d_{vc}\cdot\vec
E(s)\vec d_{cv}\cdot\vec
E(t')e^{i\omega_{ex}[k(n+1)(r-s)+t']}dt'drds],
\end{eqnarray}
where
$h_i(n)=\frac{e^{(-1)^ii\omega_{ex}k(n+i)(t-t_0)}}{\omega_{ex}k(n+i)}$
and $f_q(n)=\sqrt{\frac{q_n-q^{-n}}{q-q^{-1}}}$. These equation
shows that in this conditions second order response function is
equal zero. Now we can calculate linear and nonlinear electric
susceptibility of this exciton system from this equation.
Generalized linear and nonlinear absorption spectra of this system
is shown in figures (\ref{4})-(\ref{6}) for different values
of $q$-parameter. In these plots, $1s$-exciton is considered. In these
figures we choose $\hbar=e=1$, $g=200\;\mu ev$ and $\omega_{ex}=1574
mev$. Fig.(\ref{4}) shows plots of linear absorption spectra and
Fig.(\ref{6}) shows plots of nonlinear spectra. On the other hand,
3-dimension plot of linear absorption coefficients is given in figure
(\ref{5}). It is clear that changes of $q$-parameter
strongly affects absorption spectra of the system. These figures
show in the presence of $q$-values absorption of probe beam shows
a complex structure: a multiple-like absorption pattern appears
with one strong peak and some side bands. Presence of these side
bands is a signature of the optical generation of an nonlinear
exciton (an exciton which expresses with $q$-deformed operator).
Negative part of the absorption spectrum demonstrates gain of the
probe beam. Due to the resonance interaction of pump with exciton
transition, the gain effect comes from the coherent energy
exchange between the pump and probe beams through the QD
nonlinearity. The obtained absorption spectra are very similar to
experimental results \cite{xu}. In Ref. \cite{xu} absorption
spectra of a driven charged QD is derived experimentally. Charged
QD is a nonlinear medium and is similar to our model. Then It can
be consider as a experimental test of our model.


\section{Conclusion} $Q$-deformed description of excitons in a QD
and its physical consequences was considered. We showed that
increasing the $q$-parameter will lead to increase of splitting between peaks in the spectrum and asymmetry of spectrum. Similar effects were
observe when biexciton effects taken into account. In experiments
of QD it is shown \cite{peter} the same results are
obtained in different temperatures. Then we can associate this physical
parameter as source of $q$-deformation. The temperature dependence of emission energy
 of system can be attributed to the change in the refractive index of its active medium with
 temperature. We have derived the optical response of QD with $q$-deformed
exciton. As mentioned before $q$-deformed description of excitons
will lead to dependence of optical response on $q$ parameter.
Hence, due to the wide range of $q$ parameter and its effects on
optical response we can consider some parameters like temperature
and interaction between the excitons which affects the optical
response of QD as sources of $q$-deformation of excitons. As
mentioned, the relation of quantum statistics of excitons in the QD 
and the size of QD has been considered. Then we can consider the ratio of exciton
Bohr radius to dimension of system and exciton population as two
main sources of $q$-deformation. $Q$-deformed operator depends on
total number of associated particles of system. Therefore we can
interpret $q$-deformed operator as an operator which consists
of effects of other excitations of system implicitly. Then it is
reasonable to consider this description as an effective
description which takes into account some nonlinearity in exciton
system. As we saw, in the case of interaction of light with two
excitons, when $q=1$ this system showed a two peaks spectrum.
While by increasing deviation of exciton from Bose statistics,
spectrum becomes multi peak. Due to the nonlinear nature of
$q$-deformed exciton we showed that different orders of nonlinear
response function of this system can be calculated. From
coincidence of obtained results and experimental results, we can
conclude that $q$-deformed description of excitons can be a
considerable model for excitons. With comparing the obtained
results in this paper with experimental ones we can investigate
the origin of this description of excitons. As pointed out the
ratio of system dimension to the Bohr radius of exciton is one of
the sources of deviation of excitons from usual boson. The
obtained results are very similar to the effects of the
exciton-exciton interaction \cite{yamamoto1},\cite{moli} which is
relates to exciton population and biexciton binding energy. On the other hand, it is shown
that \cite{davydov} exciton density is another source of their
deviation from ordinary bosons. To sum up we attribute the origin
of $q$-deformation of the excitons to their density, their mutual
interactions, confinement size and other parameters which cause
fluctuation of optical response of the system.\\ $Q$-deformed description of an active medium causes that the optical properties of system depend on the $q$-parameter. Then, it is seem that parameters which can affect optical properties of the active medium (like refractive index) their effects can be considered by this formulation. The $q$-parameter can be considered as a variation parameter which its values can be obtained from comparison of theoretical and experimental results.

\textbf{Acknowledgment} The authors wish to thank
      the Office of Graduate Studies of the University of Isfahan and
      Iranian Nanotechnology initiative for
      their support.

{}
\newpage

\end{document}